\documentclass[twocolumn]{aastex62}
\usepackage{soul}

\newcommand\aastex{AAS\TeX}

\newcommand{\ie}{{\it i.e.}}
\newcommand{\eg}{{\it e.g.}}

\newcommand{\goes}{{\it GOES}}
\newcommand{\sdo}{{\it SDO}}
\newcommand{\rhessi}{{\it RHESSI}}
\newcommand{\secchi}{{\it SECCHI}}
\newcommand{\euvi}{{\it EUVI}}

\newcommand{\asecs}{\mbox{\ensuremath{^{\prime\prime}}}}
\shorttitle{\aastex\ sample article}
\shortauthors{Hernandez-Perez et al.}

\begin{document}
\title{PRE-ERUPTION PROCESSES: HEATING, PARTICLE ACCELERATION AND THE FORMATION OF A HOT CHANNEL BEFORE THE 2012~OCTOBER~20 M9.0 LIMB FLARE}

\correspondingauthor{Aaron Hernandez-Perez}
\email{aaron.hernandez-perez@uni-graz.at}

\author{Aaron Hernandez-Perez}
\affil{University of Graz, Institute of Physics/IGAM, A-8010 Graz, Austria}

\author{Yang Su}
\affil{Key Laboratory of Dark Matter and Space Astronomy, Purple Mountain Observatory\\
Chinese Academy of Sciences, 8 Yuanhua Road, Nanjing 210034, China}

\author{Astrid M. Veronig}
\affil{University of Graz, Institute of Physics/IGAM, A-8010 Graz, Austria}
\affil{Kanzelh\"ohe Observatory for Solar and Environmental Research, University of Graz, Austria}

\author{Julia Thalmann}
\affil{University of Graz, Institute of Physics/IGAM, A-8010 Graz, Austria}

\author{Peter G\"om\"ory}
\affil{Astronomical Institute, Slovak Academy of Sciences, 05960 Tatransk\'a Lomnica, Slovakia}

\author{Bhuwan Joshi}
\affil{Udaipur Solar Observatory, Physical Research Laboratory, Udaipur 313 001, India}

\begin{abstract}

We report a detailed study of the pre-eruption activities that led to the occurrence of an M9.0 flare/CME event on 2012 October 20 in NOAA AR 11598. This includes the study of the preceding confined C2.4 flare that occurred on the same AR $\sim$25~minutes earlier. We observed that the M9.0 flare occurred as a consequence of two distinct triggering events well separated  in time. The first triggering episode occurred as early as $\sim$20~minutes before the onset of the M9.0 flare, evidenced by the destabilization and rise of a pre-existing filament to a new position of equilibrium at a higher coronal altitude during the decay phase of the C2.4 flare. This brought the system to a magnetic configuration where the establishment of the second triggering event was favorable. The second triggering episode occurred $\sim$17~minutes later, during the early phase of the M9.0 flare, evidenced by the further rise of the filament and successful ejection. The second trigger is followed by a flare precursor phase, characterized by non-thermal emission and the sequential formation of a hot channel as shown by the SDO/AIA DEM (differential emission measure) maps, the \rhessi\ X-ray images and spectra. These observations are suggestive of magnetic reconnection and particle acceleration that can explain the precursor phase and can be directly related to the formation of the hot channel. We discuss on the triggering mechanisms, their implications during the early and precursor phases and highlight the importance of early activities and preceding small confined flares to understand  the initiation of large eruptive flares. 

\end{abstract}


\section{Introduction}
     \label{S-Introduction}  

Solar flares are powerful magnetically-driven events on the Sun that lead to the release of large amounts of energy in the form of particle acceleration and electromagnetic radiation \citep[\eg][]{2011SSRv..159...19F}. They are often associated with coronal mass ejections (CMEs) and affect space weather conditions \citep[for reviews see][]{2002A&ARv..10..313P,2011SSRv..159...19F,2017LRSP...14....2B}. Solar flares can be triggered by the destabilization of filaments, \ie\ dense clouds of relatively cool plasma ($\sim$8000~K), sustained against gravity by dipped or twisted (helical) magnetic fields above the Sun's surface. The latter helical structures, that are formed when the magnetic field winds about a central axis, are termed as (magnetic) flux ropes \citep[\eg,][]{2017ScChE..60.1383C}.

The pre-CME magnetic structure in the solar atmosphere is still an open issue. A flux rope may exist in the solar atmosphere already prior to the flare/CME onset, or it may be formed during the eruption \citep[for reviews see, \eg,][]{2000JGR...10523153F,2011LRSP....8....1C,2014ApJ...792L..40S,2017ScChE..60.1383C}. Flux rope formation during eruption is expected from the standard model of eruptive flares \citep{1964NASSP..50..451C,1996ASPC...95...42S,1974SoPh...34..323H,1976SoPh...50...85K}, as poloidal flux is built around a rising core field (either a sheared arcade or flux rope), due to magnetic reconnection of the embedding arcade field in a thin current sheet below the core field. Pre-existing flux ropes have been suggested to play a key role in the initiation and evolution of eruptive flares \citep{2005ApJ...630L..97T,2013ApJ...763...43C}. \cite{2041-8205-732-2-L25} presented observational evidence of pre-CME flux rope formation in the low corona, associated with coherent structures of hot plasma \citep[termed ``hot flux ropes"; see also][]{2012NatCo...3E.747Z,2013ApJ...769L..25C}. \cite{2015ApJ...808..117N} provided statistical evidence that about half of all eruptive flares may involve a hot flux rope. These are observed in hot EUV channels (such as AIA 131~\AA) and soft X-rays. For a fraction of such events, confined (CME-less) flare activity was found capable of producing and sustaining a coronal flux rope prior to eruptive flares \citep[\eg,][]{2013ApJ...764..125P,2014ApJ...795....4J,2014ApJ...789...93C,2016ApJ...832..130J}.

As summarized recently by \cite{2018SSRv..214...46G} (see Table 1 therein), there are several mechanisms that can trigger a solar eruption, \ie, that are capable of destabilizing a coronal system so that it may no longer remain in a stable equilibrium. Such mechanisms include (1) tether-cutting reconnection \citep{1992LNP...399...69M, 2001ApJ...552..833M, 2010ApJ...708..314A}, in which the destabilization occurs due to reconnection below the filament initiated by the photospheric motions; and (2) magnetic breakout \citep{1998ApJ...502L.181A,2012ApJ...760...81K}, in which, destabilization occurs due to magnetic reconnection high above the filament, which reduces the tension of the overlying fields and, thus, permitting the filament ejection. The location of the enhanced coronal X-ray emission with respect to the filament channel may shed light on the triggering mechanism behind the associated mass ejection \citep[\eg][]{2006GMS...165...43M, 2007A&A...472..967C}. Precursor X-ray emission below a rising filament may be observational evidence for tether-cutting reconnection \citep[\eg][]{2001ApJ...552..833M,2009ApJ...698..632L,2014ApJ...797L..15C,2017ApJ...834...42J}. Enhanced coronal X-ray emission above the filament may be indicative for a magnetic breakout scenario \citep[\eg,][]{2000ApJ...540.1126A,2003ApJ...595L.135J,2007SoPh..242..143J,2016ApJ...820L..37C}.

Studies of the early stages of a solar flare can not only provide information of the triggering mechanism that led to the flare, but also give a more robust understanding of the physical processes that occur before the ejection and contribute to its occurrence. Relating pre-flare physical processes with the main energy release can help to achieve a holistic understanding of why and how solar flares occur. Activity during the early phase of eruptive solar flares often involves heating \citep[\eg,][]{1996SoPh..165..169F,2002SoPh..208..297V,2017ApJ...845L...1G}, the activation and rise of a filament \citep[\eg][]{2013ApJ...771....1J,2014ApJ...797L..15C}, or flows of hot plasma along coronal loops during the precursor phase \citep[\eg][]{2015ApJ...812L..19L,2017ApJ...847..124H}. Also, plasma heating as well as electron acceleration may be observed for the predominantly thermal pre-flare X-ray corona \citep[\eg,][]{2007A&A...472..967C,2011ApJ...743..195J,2015ApJ...807..101K,2016ApJ...832..130J}.

In our paper, we present a study of the physical processes that occurred prior to the eruptive M9.0 flare on 2012 October 20 (peak time: 18:14~UT), with the aim of understanding the physical processes that brought the flaring magnetic configuration to a state in which an ejection was favorable. This includes the study of the early phase of the flare and the confined C2.4 flare that occurred in the same AR. We provide direct observations of the activation of a low-lying filament during the decay phase of the C2.4 flare, as well as of the heating and elevation of the filament during the early phase of the M9.0 flare. The fact that the source region of the flares was located behind the solar limb, provided an ideal setting to study the coronal evolution with high sensitivity. We discuss the importance of the developments in the course and aftermath of the preceding confined flare that resulted in the establishment of a hot channel, the eruptive M9.0 flare and the release of a CME.


\section{Data and Methods}
     \label{S-Data and Methods}

The Atmospheric Imaging Assembly \citep[AIA;][]{2012SoPh..275...17L} on board the {\it Solar Dynamics Observatory} \citep[\sdo;][]{2012SoPh..275....3P}, observes the ultraviolet (UV) and extreme ultraviolet (EUV) emission of the solar atmosphere, with a spatial resolution of $1\farcs2$. We studied the chromosphere, transition region and the quiet corona using AIA 304 and 171~\AA\ filtergrams (sensing plasma at about 50000\,K and 0.6\,MK respectively). Coronal and hot flare plasma is studied using AIA 193~\AA\ (sensing plasma at about 1\,MK and 20\,MK) and 131~\AA\ (sensing plasma at about 0.4 and 10\,MK). All images were co-registered and co-aligned using standard IDL {\it SolarSoft} procedures.

AIA images were processed by the noise adaptive fuzzy equalization method \citep[NAFE;][]{2013ApJS..207...25D} to enhance visibility of fine structures and their changes. In this method, two free parameters, $\gamma$ and $w$, are used to control the brightness and level of enhancement of the final (processed) image, respectively. For a detailed mathematical explanation of these parameters see \cite{2013ApJS..207...25D}. For all EUV channels, we chose $\gamma=2.6$ and $w=0.25$. For the UV channels, $\gamma=2.2$ and $w=0.2$. The unprocessed filtergrams acquired in the individual AIA channels have different dynamical ranges. Thus, the scaling parameters which determines the minimum and maximum intensity values of input and output images were set differently for all AIA channels. However, they were kept constant for the whole data set at each particular wavelength. 

This event was also observed at 195~\AA\ by the {\it Extreme UltraViolet Imager} \citep[\euvi;][]{2004SPIE.5171..111W} on the {\it Sun Earth Connection Coronal and Heliospheric Investigation} \citep[\secchi;][]{2008SSRv..136...67H} instrument suite, onboard the Solar TErrestrial RElations Observatory \citep[STEREO;][]{2008SSRv..136....5K}. The event occurred on the western limb from Earth view, \ie\ was visible on-disk for STEREO-B (ST-B).

X-ray imaging and spectroscopy studies were performed with the {\it Reuven Ramaty High Energy Spectroscopic Imager} \citep[\rhessi;][]{2002SoPh..210....3L}. \rhessi\ observes X-rays and gamma-rays at energies ranging from 3~keV to 17~MeV with a time cadence of 4~seconds. It has a spatial resolution as high as $2\asecs$ and a spectral resolution of $\sim$1~keV. For the reconstruction of X-ray images in this study, we used the CLEAN algorithm \citep{2002SoPh..210...61H} using the front detectors 3--9, excluding 4. For the spectral analysis we used only data from the front segment of detector 6, as its spectrum most closely resembles the mean spectrum based on the measurements of all detectors, for the $\sim$4--100~keV energy band and an integration time of 28~s. The spectral fitting was computed using a combined model that includes one or more isothermal models and a thick-target non-thermal emission model \citep{1971SoPh...18..489B,2003ApJ...595L..97H}. 

In order to study the thermal evolution of the flare plasma, we also calculated the Differential Emission Measure (DEM) maps for the early phase of the M9.0 flare, using the Sparse inversion method \citep{2015ApJ...807..143C}, with the new settings proposed by \cite{2018ApJ...856L..17S}.

\begin{figure*}[ht]
\centerline{
\centering\includegraphics[width=\textwidth]{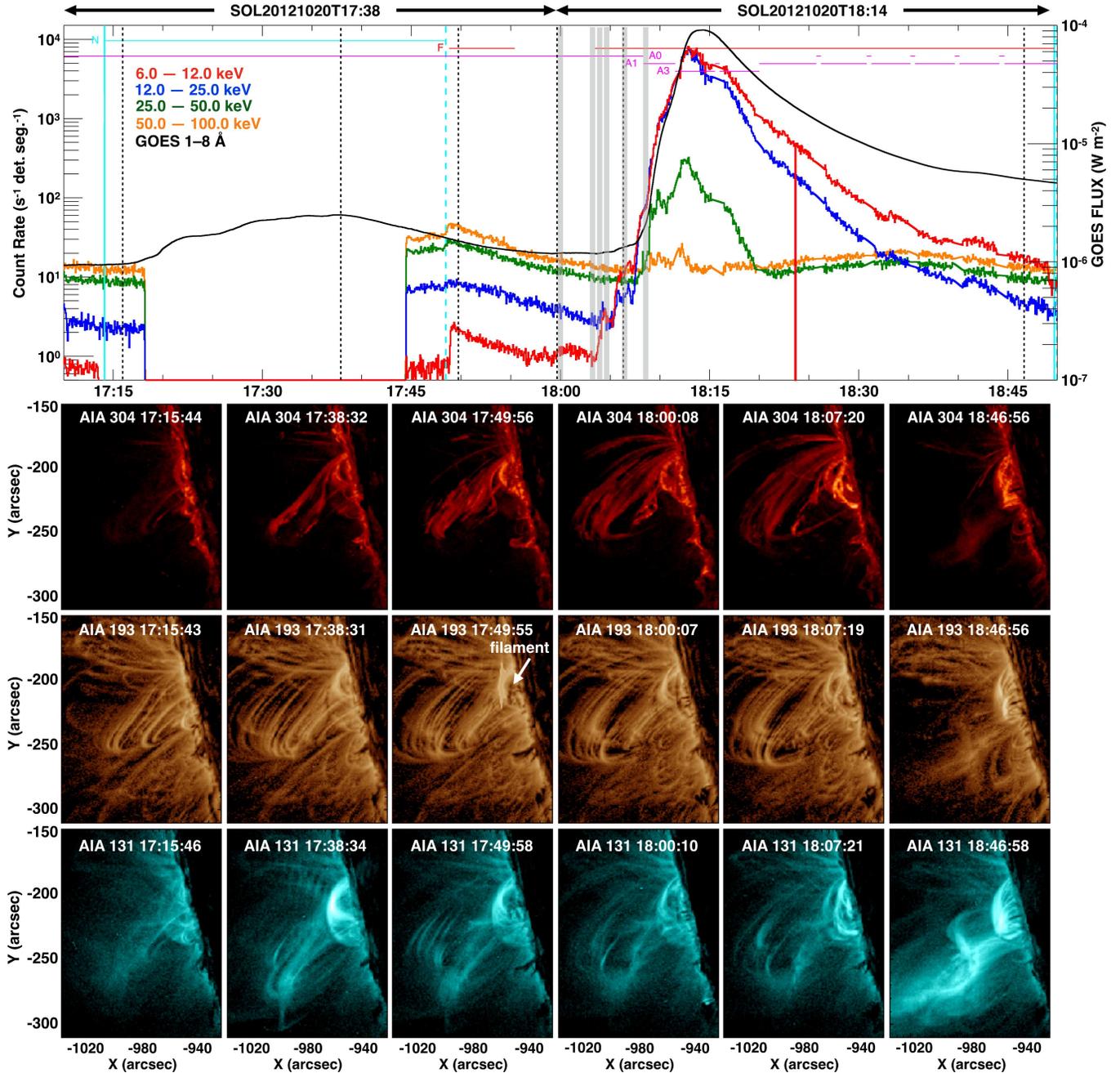}
}
\caption{
Top: \rhessi\ X-ray count rates for the 6--12 (red), 12--25 (blue), 25--50~keV (green) and 50--100~keV (orange) energy bands and the \goes\ SXR flux in the 1--8~\AA\ wavelength band (black thick solid line). The cyan, red and pink segments indicate the \rhessi\ night time (N), the flaring activity (F) and the attenuator states (A0--A3). Bottom: sequence of AIA 304~\AA\ (top panels), AIA 193~\AA\ (middle panels) and AIA 131~\AA\ (bottom panels) images. The panels show the EUV flare emission for the times indicated by black dashed lines on the top panel. The animation attached to this image show the evolution of the flare in co-temporal maps at 304, 193 and 131~\AA\ for the field of view (FOV) shown in the images.\\
(An animation of this figure is available.)
}
\label{Lightcurve}
\end{figure*}

\section{Results}
     \label{S-Results}

\subsection{Event Overview}
     \label{S-Event Overview}

\begin{figure}[ht]
\centerline{
\centering\includegraphics[width=0.5\textwidth]{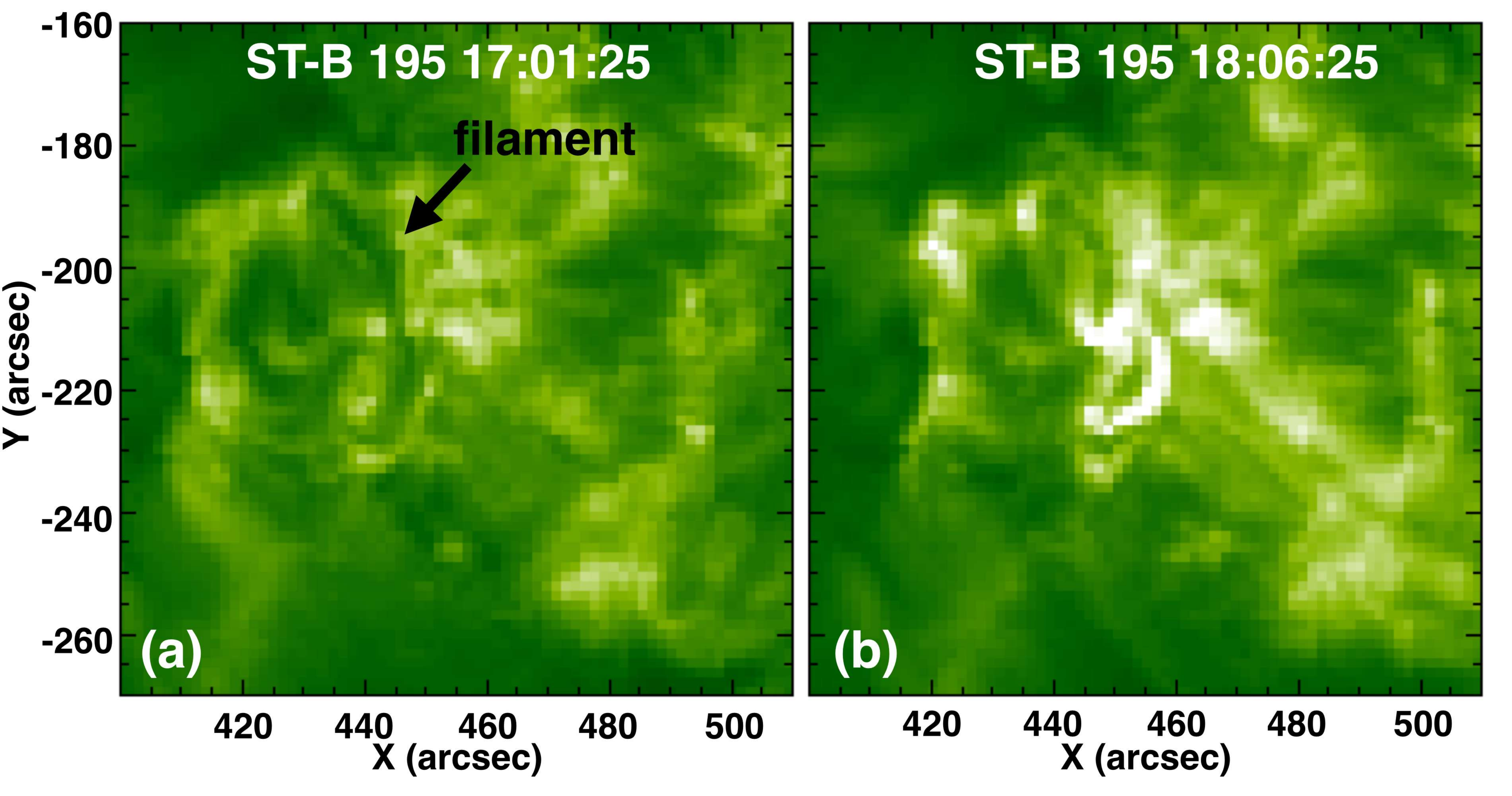}
}
\caption{
ST-B EUVI 195~\AA\ sequence of the AR site at different times, (a) before the confined C2.4 flare and (b) during the early phase of the M9.0 flare (\ie\ SOL20121020T18:14).\\
}
\label{stereo}
\end{figure}

\begin{figure*}[ht]
\centerline{
\centering\includegraphics[width=\textwidth]{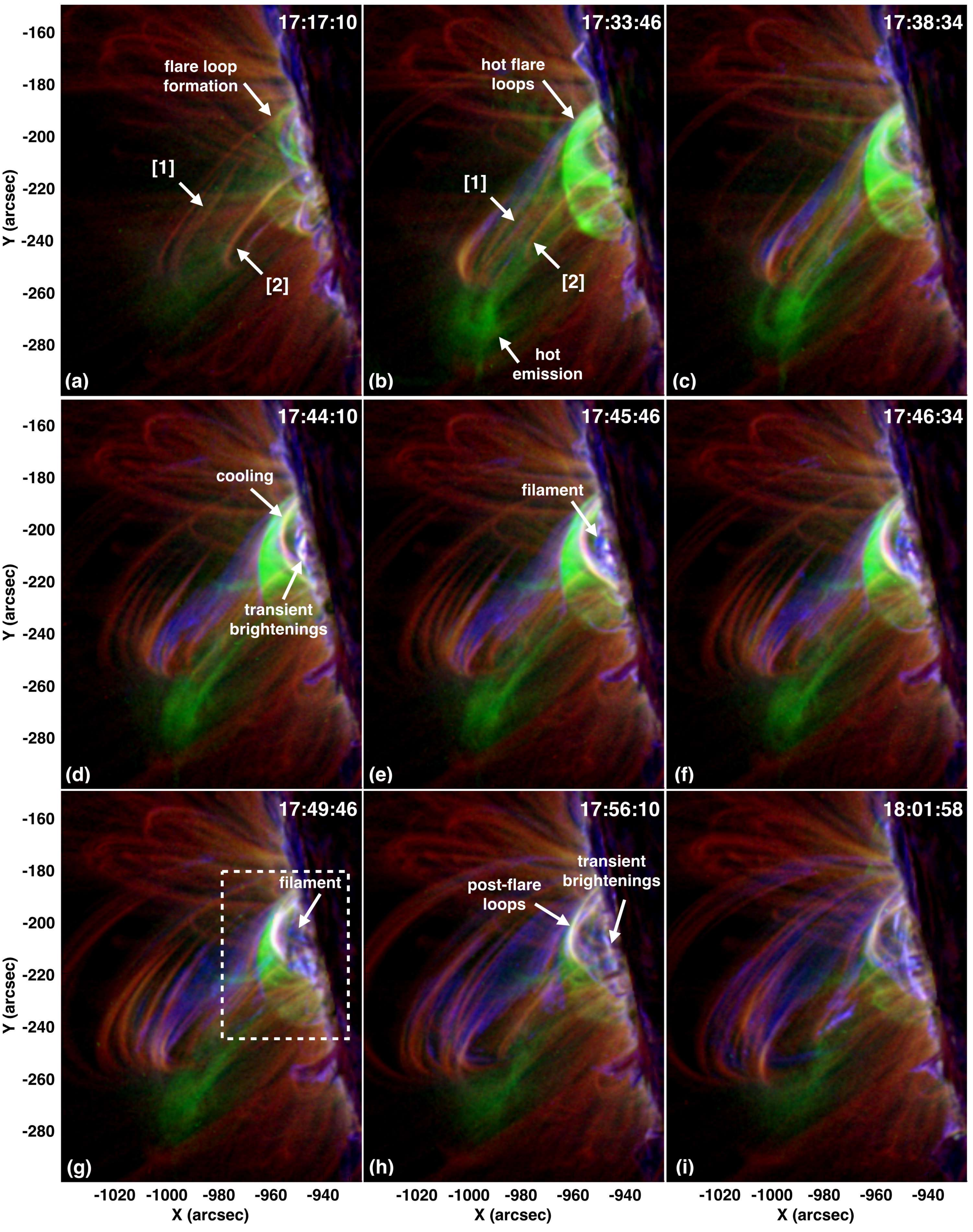}
}
\caption{
Composite sequence of images in AIA 304~\AA\ (blue), 171~\AA\ (pink) and 131~\AA\ (green) during the course of the C2.4 flare and the early phase of the M9.0 flare. The animation attached to this image shows the composite sequence for the two flares.\\
(An animation of this figure is available.)
}
\label{composite_cflare}
\end{figure*}

\begin{figure*}[ht]
\centerline{
\centering\includegraphics[width=\textwidth]{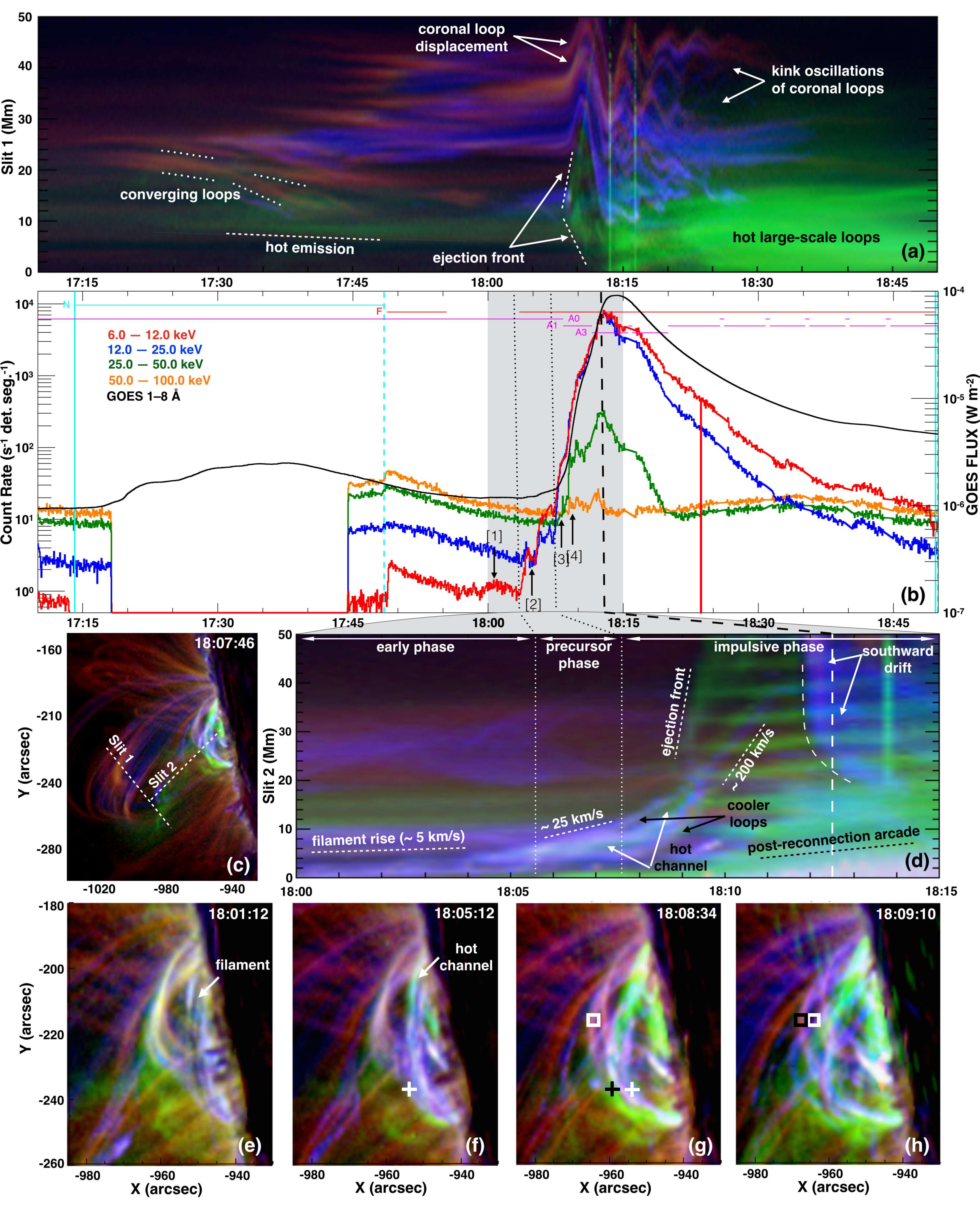}
}
\caption{
(a) Composite time-distance plot for slit 1 (see panel (c)), (b) \rhessi\ count rates and \goes\ flux (see Fig.~\ref{Lightcurve} for details). (c) Composite AIA image of the flare before the ejection, (d) Composite time-distance plot for slit 2 (see panel (c)), (e--h) zoomed composite of AIA images (area marked by a white dashed line in Fig.~\ref{composite_cflare}(g)) showing the formation and evolution of the hot channel (see times indicated by black arrows [1--4] in (b)). The colors represent AIA 304 (blue), 171 (pink) and 131~\AA\ (green).
}
\label{Stack_plots_hot_channel}
\end{figure*}

On 2012 October 20, AR NOAA 11598 produced four flares on the Eastern limb of the Sun: a C1.8 flare (\goes\ SXR peak time 14:44~UT), a C2.4 flare (peak time 17:38~UT), an M9.0 flare (peak time 18:14~UT) and a C2.7 flare (peak time 20:09~UT). All were confined events, except for the M9.0 flare, which had a slow CME associated, with a projected speed of $\sim$320~km~s$^{-1}$ \citep{2009EM&P..104..295G}. In this paper, we mainly concentrate on the activity that led to the eruptive M9.0 flare peaking at 18:14~UT (SOL20121020T18:14). This includes a detailed study of the early phase of the M9.0 flare, as well as of the preceding confined C2.4 flare (SOL20121020T17:38) and how the two flare events are related to each other.


The top panel of Fig.~\ref{Lightcurve} shows the \rhessi\ X-ray count rates at different energy bands constructed using the front segments of detectors 3, 5, 6, 7, 8, and 9, along with the  \goes\ 1--8~\AA\ SXR flux. The bottom panels show the coronal EUV emission for selected times in AIA 304, 193 and 131~\AA\ images. An animation\footnote{For quality purposes, the cadence for the time interval $\sim$18:10--18:31~UT is of 24~s in animations 1 and 2.} is attached to this image showing the evolution of the flare in co-temporal maps at 304, 193 and 131~\AA\ for the field of view (FOV) represented in the figure (animation 1).

\rhessi\ data covered part of the decay phase of the confined C2.4 flare and the full evolution of the eruptive M9.0 flare. Note that between $\sim$17:46 and $\sim$18:08~UT, \rhessi\ observed continuously in attenuator state A0. Therefore, the data during that time span are of high sensitivity. The early phase of the M9.0 flare terminates with the sudden X-ray enhancement at 18:07:30~UT. The rest of the impulsive phase is characterized by two HXR bursts in the 25--50 and 50--100~keV energy bands peaking at around 18:10 and 18:12~UT (see Fig.~\ref{Lightcurve}).

AIA images reveal the presence of a dark filament (see white arrow in Fig~\ref{Lightcurve}). Prior observations to the C2.4 flare do not show signatures of this filament. Therefore, in order to find out whether or not the filament was present prior to the C2.4 flare, we have explored STEREO-B (ST-B) EUVI observations. On that day, ST-B was separated by 126$^{\circ}$ in longitude from Earth, thus viewing the event on disk. Fig.~\ref{stereo} shows on-disk images of the AR by ST-B at 195~\AA. It can be observed that the filament was already present prior to the confined C2.4 flare. Further observations also showed that it was also present prior to the C1.8 flare observed first in the series of events. 

For a thorough characterization of the initiation and early phases of the eruptive M9.0 flare, and how it leads to its impulsive phase, we separate the description in three parts. In Section \ref{S-The C2.4 flare and flux rope activation} we study the confined C2.4 flare and present observations of the most plausible mechanism responsible for the initiation of the rise of the filament. In Section \ref{S-The M9.0 flare: Early phase} we study the early phase of the M9.0 flare, and give observational evidence of a slow quasi-static rise of the filament preceded by a precursor phase, signatures of particle acceleration and the formation of a hot channel. In Section \ref{S-The M9.0 flare: Impulsive and Main phases} we present a description of the impulsive phase of the M9.0 flareand the ejection of the filament. In Section \ref{S-Flux Rope heating}, for completeness, we focus on the heating characteristicsduring the filament activation.

\subsection{The C2.4 flare and initiation of the M9.0 flare}
     \label{S-The C2.4 flare and flux rope activation}

\begin{figure*}[ht]
\centerline{
\centering\includegraphics[width=\textwidth]{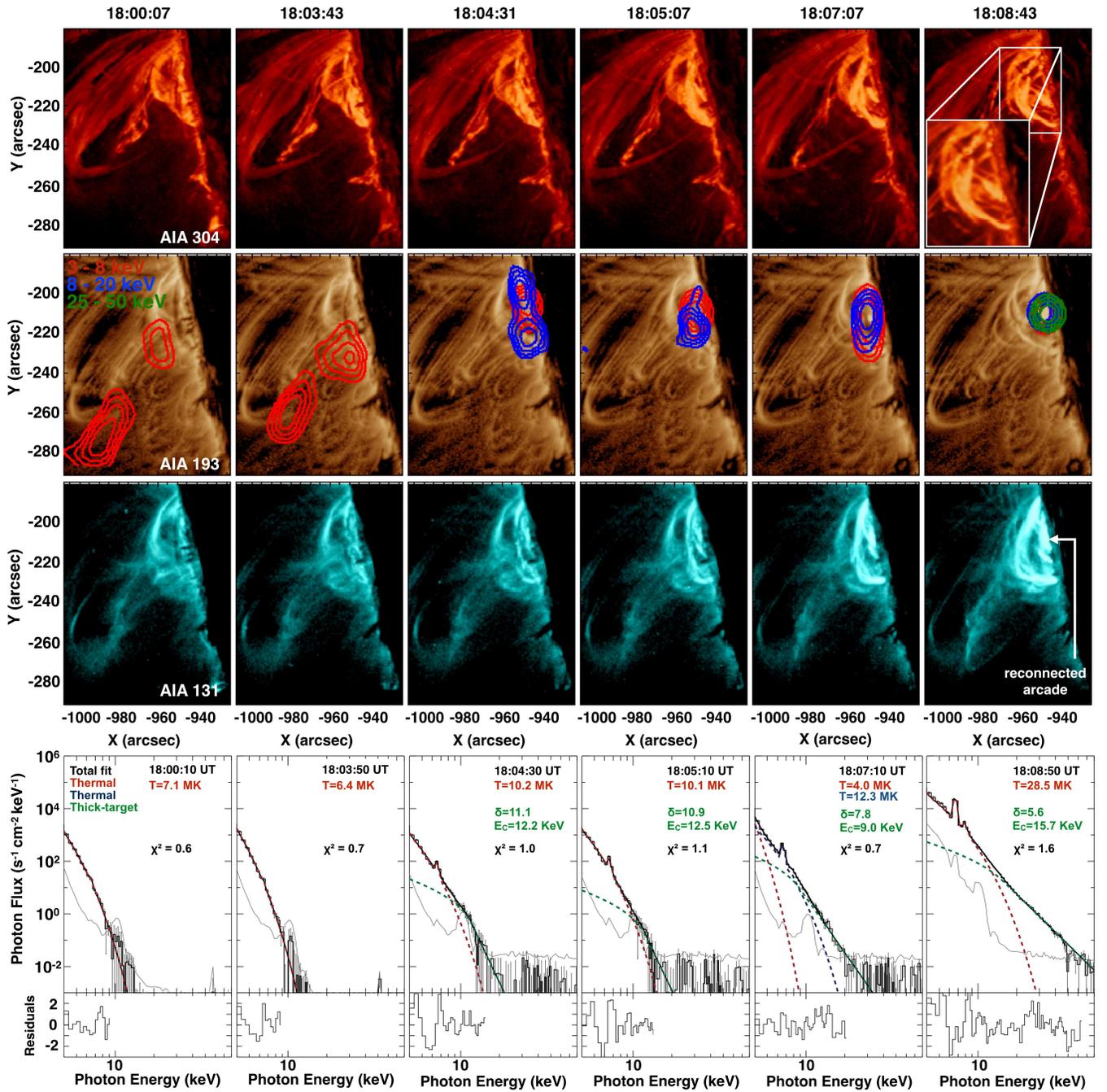}
}
\caption{
Top: AIA image sequence at 304, 193 and 131~\AA\ during the early phase of the M9.0 flare for the low corona. On top of the 193~\AA\ sequence, the \rhessi\ images for 3--8 (red), 8--20 (blue) and 25--50~keV (green) are overplotted for 60, 70, 80 and 90\% of the maximum emission. Bottom: corresponding \rhessi\ spectra for the time sequence (see time intervals marked by shaded areas on the top panel of Fig.~\ref{Lightcurve}). For direct comparison with animations 1 and 2, the times displayed at the top correspond to the AIA 193~\AA\ images.}
\label{Spectra}
\end{figure*}
 
\begin{figure*}[ht]
\centerline{
\centering\includegraphics[width=\textwidth]{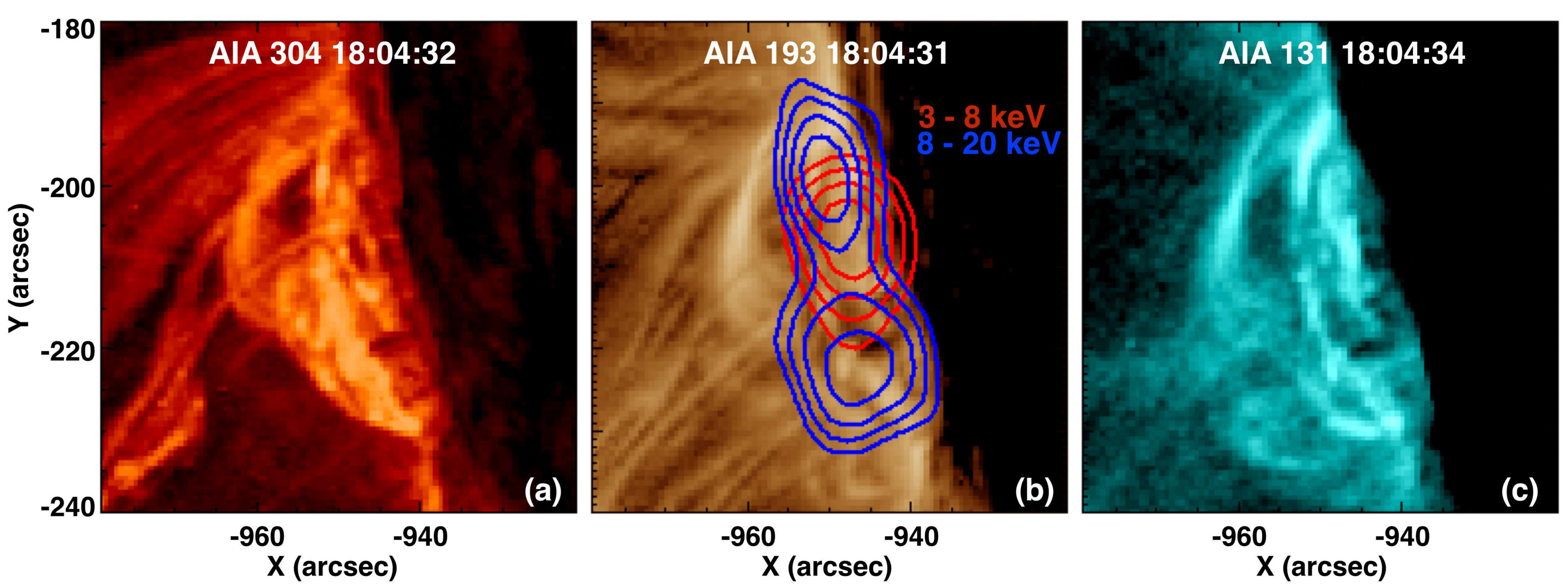}
}
\caption{
Zoomed AIA images at 304, 193 and 131~\AA\ at the time of the first precursor (\ie\ $\sim$18:04:30). The \rhessi\ images for 3--8 (red) and 8--20 (blue) are overplotted for 60, 70, 80 and 90\% of the maximum emission.}
\label{Precursor}
\end{figure*}

Fig.~\ref{composite_cflare} shows a composite sequence of images in AIA 304~\AA\ (blue), 171~\AA\ (pink) and 131~\AA\ (green) during the course of the C2.4 flare and the early phase of the M9.0 flare. An animation is attached to this figure (animation 2) showing the composite sequence for the two flares. During the course of the confined C2.4 flare, a flare loop arcade is observed to form (see Fig.~\ref{composite_cflare}(a,b)). Overlying the forming loop arcade, bundles of large-scale coronal loops (labeled ``[1]" and ``[2]") are observed to approach each other (best observed in AIA 171~\AA). This converging motion of the loops is co-temporal with an increase of hot emission high in the corona as observed in AIA 131~\AA\ (compare Fig.~\ref{composite_cflare}(a,b)). This suggests localized heating, most likely due to reconnection, high in the corona. Fig.~\ref{Stack_plots_hot_channel}(a) shows a composite time-distance plot during the two events for slit 1 (represented by the white dashed line in Fig.~\ref{Stack_plots_hot_channel}(c)), for comparison, Fig.~\ref{Stack_plots_hot_channel}(b) shows the \rhessi\ count rates and \goes\ flux as depicted in Fig.~\ref{Lightcurve}. The time-distance plot reveals that loops overlying the post-reconnected arcades converge (see cool converging loops observed in AIA 304 and 171~\AA\ marked with white dashed lines in Fig.~\ref{Stack_plots_hot_channel}(a)) toward a hot emission region (in green) indicative of heating. This suggests ongoing reconnection high above the post-flare loops of the C2.4 flare.

During the decay phase of the C2.4 flare, the post-flare loops are observed to cool down as is \eg\ evidenced by the decaying hot 131~\AA\ emission and cool 193 and 171~\AA\ emission (see Fig.~\ref{composite_cflare}(d--i) and animation 1). This cooling is co-temporal with the occurrence of transient brightenings and structural changes, as well as the rise of the filament $\sim$20~minutes before its ejection (see Fig.~\ref{composite_cflare}(d,e) and animation 2). This is followed by a slow rise of the filament and its temporary stabilization at a higher altitude (see Fig.~\ref{composite_cflare}(f,g)). The signatures of heating in the higher corona remain present during the ascent of the filament, indicative of ongoing reconnection overlying the filament (see also Fig.~\ref{Stack_plots_hot_channel}(a)). The filament is maintained at a stable height for $\sim$10~minutesthereby becoming part of the early phase of the M9.0 flare, where transient brightenings are observed to occur beneath the filament while the filament is at this new (higher) equilibrium (see Fig.~\ref{composite_cflare}(h,i)).

\subsection{Early phase of the M9.0 flare}
     \label{S-The M9.0 flare: Early phase}

Fig.~\ref{Stack_plots_hot_channel}(d) shows a composite time-distance plot for slit 2 (see slit 2 represented in Fig.~\ref{Stack_plots_hot_channel}(c)) for the early and impulsive phases of the M9.0 flare (see shaded area in Fig.~\ref{Stack_plots_hot_channel}(b)). During the early phase of the M9.0 flare, the filament can be observed to undergo a slow quasi-static rise with a projected velocity of $\sim$5~km/s (see Fig.~\ref{Stack_plots_hot_channel}(d) and animations 1 and 2). Transient brightenings that occur below the filament during the rise are also observed, indicating that the corona is undergoing small-scale structural changes (for comparison see also Fig.~\ref{composite_cflare}(i)).

Subsequent to this phase of slow rise, an acceleration of the elevating structure takes place, and a faster rise with a projected velocity of $\sim$25~km/s is observed, co-temporal with a precursor phase, \ie, SXR activity in the form of two distinct episodes of flux enhancement in the 6--12 and 12--25~keV \rhessi\ count rates (see the precursor phase delimited by the dotted lines in Fig.~\ref{Stack_plots_hot_channel}(b,d)). In order to study the X-ray emission during the early and impulsive phases, Fig.~\ref{Spectra} shows a time sequence of AIA 304, 193 and 131~\AA\ images and their corresponding \rhessi\ spectra. \rhessi\ contours levels at 60, 70, 80 and 90\% of the maximum emission are plotted for 3--8 (red), 8--20 (blue) and 20--50~keV (green) energy bands on top of the AIA 193~\AA\ images.

Two thermal X-ray sources in the energy range of 3--8~keV are firstly observed (see images at 18:00:07 and 18:03:43~UT in Fig.~\ref{Spectra}). This represents hot X-ray emitting plasma in coronal loops before the M9.0 flare (see also EM maps for hot plasma at 6--12~MK in Fig.~\ref{EM2}). At the first peak of the SXR precursor phase (see third column in Fig.~\ref{Spectra}), the filament brightened (best observed in AIA 131~\AA). The \rhessi\ X-ray emission at 3--8~keV is more compact and comes from below the filament. More energetic X-ray emission in the range of 8--20~keV appear, located to the north and south of the 3--8~keV source. 

For a better overview of the X-ray sources at this time, Fig.~\ref{Precursor} shows the zoomed AIA images at 304, 193 and 131~\AA\ at the time of the first precursor. The 8--20~keV emission originates close to the observable ends of the filament, where it starts to brighten (see corresponding image at 131~\AA). AIA 1600~\AA\ observations do not show signatures of foot points at this time and therefore they are occulted behind the limb. Hence, we identify the emission in the 8--20~keV energy range to originate from the legs of the filament. The spectral analysis reveals the thermal nature of the 3--8~keV source with T $\sim$10~MK, and also the existence of non-thermal emission at energies $\gtrsim$12~keV, indicative of plasma heating and particle acceleration during the early phase (see Fig.~\ref{Spectra}).

The subsequent X-ray emission during the early phase is characterized by two compact sources located below the filament with the exception of the 8--20~keV source at the time of the second precursor (\ie\ 18:05:07~UT, see Fig.~\ref{Spectra}), found to be located at the the southern leg of the elevating filament, which has become brighter at this time (compare Fig.~\ref{stereo}(d) with the fifth AIA 131~\AA\ image in Fig.~\ref{Spectra}) indicative of heating. The twisted morphology of the structure comprising the elevating filament starts to become observable at this time (see the inset in Fig.~\ref{Spectra}). This twisted nature indicates that this structure may be a flux rope \citep[\eg][]{2013ApJ...763...43C}. This will be discussed in Sect.~\ref{S-Flux Rope heating}. The strongest emission of the X-ray sources is located beneath the rising filament, where a newly reconnected arcade is being formed (see Fig.~\ref{Spectra}). The spectra show that the non-thermal power-law spectra become flatter, due to electrons accelerated to higher energies. The temperature of the plasma as measured from the thermal fitting is also observed to increase.

\subsection{Ejection and impulsive phase of the M9.0 flare}
     \label{S-The M9.0 flare: Impulsive and Main phases}

At the beginning of the impulsive phase (see images at 18:08:43~UT in Fig.~\ref{Spectra}), the filament has considerably risen and a newly detected HXR source in the 20--50~keV energy range appears, co-spatial with those at 3--8 and 8--20~keV. The maximum emission of the X-ray sources comes from the location where the reconnected arcade is being formed (compare images at 193 and 131~\AA\ at 18:08:43~UT in Fig.~\ref{Spectra}). Further X-ray observations reveal emission in the 3--50~keV energy range coming from this location until the end of the impulsive phase (\ie\ $\sim$18:15~UT).

We emphasize that the fast rise of the filament is co-temporal with a steep increase of the HXR (25--50~keV) emission (see Fig.~\ref{Stack_plots_hot_channel}(b,c)). The projected velocity of the ejection during the impulsive phase is $\sim$200~km~s$^{-1}$. The overarching coronal loops interact with the erupting structure (see eruption front in Fig.~\ref{Stack_plots_hot_channel}(a,d)), and the overarching fields are displaced in northward direction (see animation 2). A post-reconnection loop arcade becomes visible in the low corona co-spatial with HXR emission. After the ejection, the pre-flare configuration of the overarching loops is restored and they are observed to oscillate (see Fig.~\ref{Stack_plots_hot_channel}(a)). These oscillations have been identified as \textit{kink oscillations} \citep{1999Sci...285..862N,2016A&A...585A.137G}. The HXR curve undergoes further rise peaking at $\sim$18:12:30~UT, which coincides with the epoch when the large-scale loops reach their minimum height (see dashed lines in Fig.~\ref{Stack_plots_hot_channel}(b,d)). Fig.~\ref{Lightcurve}(h) shows the late magnetic structure left by this flare, the post-flare loops are visible in the low corona together with the very hot overarching coronal loops visible in hotter channels (see animation 2 for a composite view of the decay phase of the M9.0 flare).

\subsection{Analysis of filament activation and hot channel formation}
      \label{S-Flux Rope heating} 

Motivated by the fact that the filament underwent a pre-heating phase before its ejection (see last image in AIA 131 in Fig.~\ref{Spectra}), we present in this section EUV and DEM analyses using the AIA EUV images of the early phase of the event with the aim of determining precisely when and how the heating of the filament occurred.

Fig.~\ref{Stack_plots_hot_channel}(e--h) shows a sequence of composite images of the solar corona, as observed in the three wavebands 304~\AA\ (blue), 171~\AA\ (pink) and 131~\AA\ (green). Corresponding times are indicated by black arrows [1--4] in Fig.~\ref{Stack_plots_hot_channel}(b). These still images are representative of the early and impulsive phases. The sequence reveals the formation and evolution of a hot channel (in green), in agreement with the findings in \cite{2012NatCo...3E.747Z}. The sequential heating suggests that the hot channel was formed as a consequence of the heating of a pre-existing structure within the flux rope. The black and white crosses represent the apex of the hot channel at 18:05:12~UT and 18:08:34~UT respectively. The black and white squares represent the apex of the post-flare loop left by the C2.4 flare at 18:08:34~UT and 18:00:10~UT respectively.

Based on the comparison of Fig.~\ref{Stack_plots_hot_channel}(e,f), the hot channel must have formed around 18:05~UT, \ie, $\sim$2~minutes before the impulsive onset of the M9.0 flare. In the subsequent intervals (\ie\ during the precursor phase and early impulsive phase), the hot filament channel rose in altitude (by $\sim$5~arcsec; compare the black and white crosses in Fig.~\ref{Stack_plots_hot_channel}(f,g) marking the apparent apex of the flux rope structure). Noteworthy is the rise of the post-flare loops left by the C2.4 flare, which also rose in altitude with the rising hot channel (by $\sim$3~arcsec; compare the black and white squares in Fig.~\ref{Stack_plots_hot_channel}(g,h)). Animation 2 shows the interaction of the overlying post-flare loops left by the C2.4 flare and the elevating structure, co-temporal with the sudden increase in the \rhessi\ HXRs (compare with Fig.~\ref{Stack_plots_hot_channel}(b)).

\begin{figure*}[ht]
\centerline{
\centering\includegraphics[width=\textwidth]{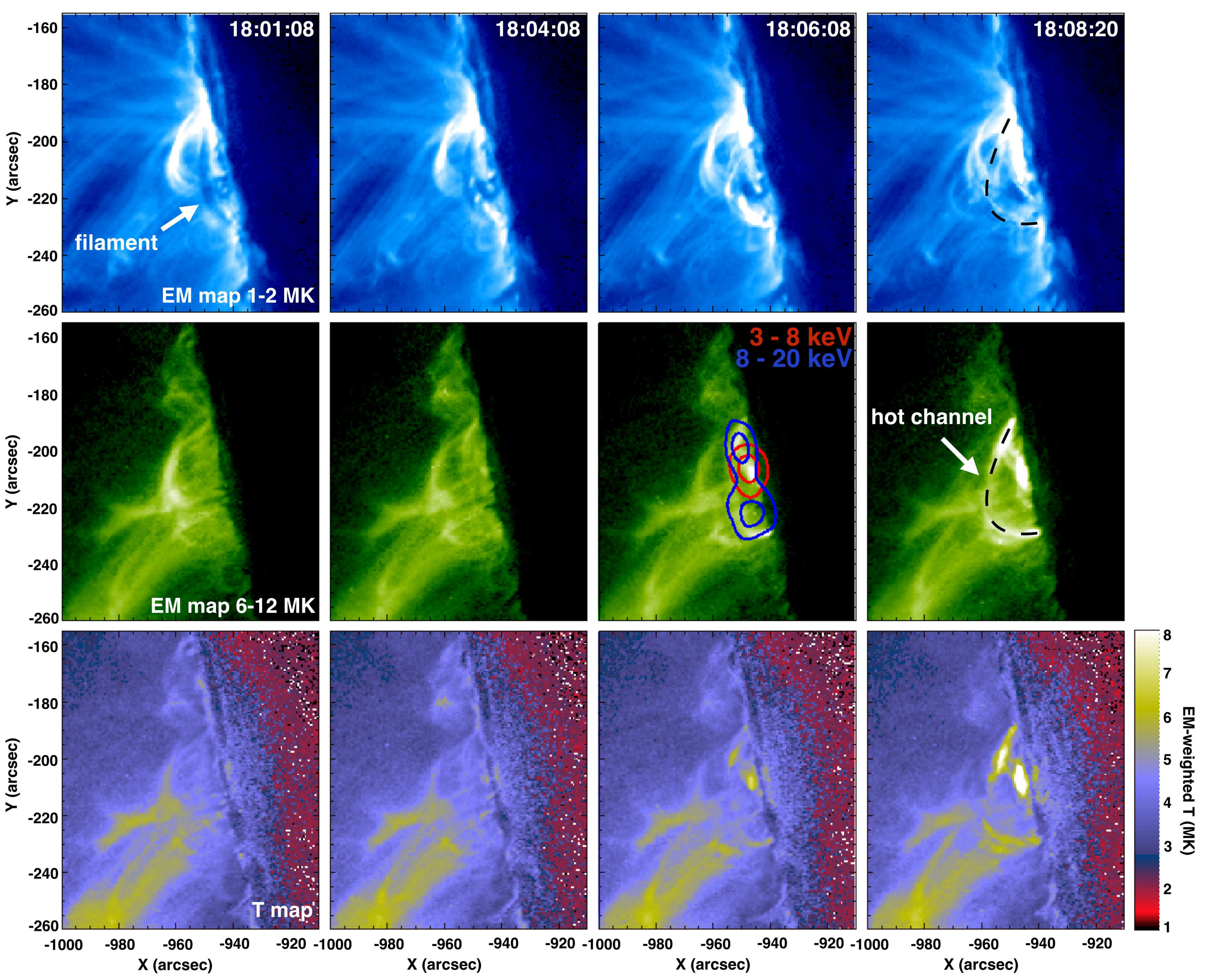}
}
\caption{
EM maps at 1--2~MK (top) and at 6--12~MK (middle) and temperature maps (bottom) during the early phase of the M9.0 flare. Plotted over the 6--12~MK EM map at 18:06:08~UT are the \rhessi\ sources for 3--8~keV (in red) and for 8--20~keV (in blue) at 70 and 90\% of the maximum emission at the time of the first SXR precursor (\ie\ at 18:04:30~UT, see Fig.~\ref{Spectra}). The dashed line plotted in the last 6--12~MK EM map represents the hot core.\\
(An animation of this figure is available.)
}
\label{EM2}
\end{figure*}

Exploration of the formation mechanisms of hot channels is fundamental in the understanding of the initiation processes of CMEs and flares. To understand the formation of the hot flux rope in detail, we performed a DEM analysis and present EM and weighted temperature maps in Fig.~\ref{EM2}, covering the early and early impulsive phases of the M9.0 flare. The EM maps are shown for two temperature ranges, 1--2~MK (top row panels) and at 6--12~MK (second row panels). \rhessi\ X-ray emission at the time of the first SXR precursor (\ie\ $\sim$18:04:30~UT) is displayed on top of the 6--12~MK EM map at 18:06:08~UT. Contours are drawn for the 3--8~keV (in red) and for 8--20~keV (in blue) energy band, at 70 and 90\% of the maximum emission. The attached on-line animation (animation 3) clearly depicts the heating of the filament during the early flare phase. At the time when the two non-thermal X-ray sources appeared co-spatial with the legs of the flux rope (\ie\ $\sim$18:04~UT), the 6--12~MK EM map does not show significant enhancement from those locations. This indicates slow heating at the time where the non-thermal sources were found.

At $\sim$18:06:08~UT the filament started to rise and heat. The weighted T-maps reveal that first the legs of the filament channel heat up and subsequently its apex (compare weighted T-maps at 6--12~MK at 18:06:08 and 18:08:20~UT).

\section{Discussion and Conclusion} 
      \label{S-Discussion and Conclusion} 

We present detailed analyses of flare activity observed above the solar limb essential to understand the activation and ejection of a pre-existing filament, which led to an eruptive M9.0 flare (SOL20121020T18:14) in NOAA~11598. The pre-CME activities include plasma heating and particle acceleration, as well as plasma motions in and around the pre-existing filament. Noteworthy, we were also able to image the sequential formation of an associated hot channnel, indicative of a hot flux rope.

The first distinct physical signature of early activity, that set a favorable configuration for an ejection, occurred during the decay phase of the preceding confined C2.4 flare, $\sim$20~minutes before the eruption. This was the rise of the low-lying filament to a higher position of equilibrium (see Fig.~\ref{composite_cflare}(d--i)). The phase of activation and ascent of the filament during which it moves to a stable height prior to its subsequent successful eruption has been observed in earlier studies \citep[\eg][]{2013ApJ...764..125P,2013ApJ...771....1J,2014ApJ...797L..15C,2016ApJ...832..130J}. High above the filament, we evidenced signatures for coronal magnetic reconnection in the form of heating above the filament and in addition, converging motions of coronal loops toward the probable reconnection region (Fig.~\ref{Stack_plots_hot_channel}(a)). This reconnection of high coronal loops above the filament results in the reorganization of coronal fields and reduces the tension of the overlying loops, affecting the force equilibrium of the low-lying magnetic structure and facilitating its slow expansion, therefore causing its destabilization \citep{2016ApJ...820L..37C}. Furthermore, SXR emission in the range of 3--8~keV, was found to be co-spatial with the hot emission produced during the course of the C2.4 flare (see Fig.~\ref{Spectra}), representative of hot X-ray emitting plasma in coronal loops. 

During the early phase of the eruptive M9.0 flare (see Fig.~\ref{Stack_plots_hot_channel}(d)), we observed a slow quasi-static rise of the filament with a projected speed of $\sim$5~km/s, at the time of its activation, co-temporal with EUV brightenings below the filament (compare with Fig.~\ref{composite_cflare}(h,i)), indicative of resistive processes. SXR precursors, indicative of the time scale and strength of episodic small-scale energy release (similar to what was reported by, \eg, \cite{2009ApJ...698..632L} (see Fig.1a therein)), were observed co-temporal with an increase in the projected velocity of the rise of the filament (compare Fig.~\ref{Stack_plots_hot_channel}(b,d)). This can be explained by the slow expansion of the filament, that causes internal reconnection, which also promotes the further expansion of the system \citep{2001JGR...10625227S,2016ApJ...832..130J}. In fact, the synthesis of X-ray and EUV observations during the precursor phase (see Fig.~\ref{Spectra} and \ref{Precursor}) provide clear evidence that magnetic reconnection occurred within the core region of the flaring environment, below the apex of a quasi-stationary low-lying flux rope as well as close to its two end points (Fig.~\ref{Precursor}). This points toward the small-scale localized magnetic reconnection, taking place in the core fields \citep[][]{1992LNP...399...69M,2001ApJ...552..833M}. The slight yet distinct non-thermal component of the X-ray emission observed during this phase implies the acceleration of particles resulting from these reconnection events. 

The slow quasi-static rise of the filament was followed by an increase in the upward velocity of the filament (see Fig.~\ref{Stack_plots_hot_channel}(d)), indicating that an additional triggering episode took place, further destabilizing the system and causing its ejection. The X-ray sources detected below the ascending flux rope, co-spatial with the locations where the low-lying post-reconnected loops formed, are indicative of resistive processes at this location. The second triggering episode could therefore be explained as a combination of (1) an ideal instability, supported by the continuous quasi-static rise of the magnetic structure, and (2) tether-cutting reconnection, supported by the observation of X-ray sources below the rising magnetic structure already during the precursor phase.

The fast rise of the magnetic structure is co-temporal with the steep increase in the HXR (25--50~keV) emission (compare Fig.~\ref{Stack_plots_hot_channel}(b,d)). Two processes can energize the CME, (1) the large-scale Lorentz force acting on the flux rope, and/or (2) the dissipative process of reconnection \cite{2017ScChE..60.1383C}. The outward contribution of the newly reconnected fields below the rising flux rope provide additional poloidal flux increasing the hoop force and contributing to the flux rope acceleration \citep{2016AN....337.1002V}. As shown in the simulations by \cite{2007ApJ...665.1421C}, the two processes often occur synchronously and provide positive feedback to each other. Thus, they are often not separable through present observations.

AIA DEM analysis revealed the formation of a hot channel as a sequential heating of a pre-existing loop within the flux rope. Hot channels were first reported in \citep{2012NatCo...3E.747Z}, and have been regarded as evidence of magnetic flux ropes. Our observations show two main differences with respect to their observations: 1. the earliest signature of the hot channel was observed as early as $\sim$2~minutes before the SXR onset instead of $\sim$6~minutes. This was evidenced by two localized EUV enhancements at the location of the legs of the filament co-spatial with two non-thermal sources (see Fig.~\ref{Precursor}). This shows that hot channels can be formed very close in time to their ejection; 2. its formation occurred along with the slow rise of the flux rope.The temperature of the isothermal plasma as measured from the thermal fitting is also observed to increase.

Based on these results, we conclude that 1. a first triggering episode occurred during the decay phase of the C2.4 flare as a consequence of magnetic reconnection of coronal loops overlying the filament. This reduced the tension of the overlying fields, leading to the rise of the filament to a higher position where no stable equilibrium existed \citep{1991ApJ...373..294F,1995ApJ...446..377F, 2000JGR...105.2375L}, 2. after the quasi-static rise of the filament during the early phase of the M9.0 flare, a second triggering episode occurredmost probably as a combination of an ideal instability and tether-cutting reconnection. 

In summary, we show that the major eruption occurred as a consequence of two distinct triggering episodes separated in time by $\sim$17~minutes. The first trigger occurred as early as $\sim$20~minutes before the onset, \ie\ when the filament was destabilized during the preceding confined C-class flare. This left the magnetic configuration in a state where equilibrium was no longer possible, leading to a subsequent quasi-static rise of the filament during the early phase of the M9.0 flare until a second triggering episode occurred $\sim$3~minutes before the onset. Our work highlights the importance in extending flare studies by also considering preceding (possibly confined) flare events during the early phase of eruptive flares, in order to understand the initiation of large flares and their associated  fast eruptions.

\acknowledgments
\small

The authors gratefully acknowledge  the  anonymous  referee  for  the  comments  that significantly helped to improve the manuscript. A.H., A.M.V. and J.K.T. acknowledge the Austrian Science Fund (FWF): P27292-N20. This work is supported by the Indo-Austrian joint research project no. INT/AUSTRIA/BMWF/ P-05/2017 and OeAD project no. IN 03/2017 and by the bilateral OeAD/SRDA Austrian-Slovak project no. SK-AT-2017-0009. P.G. acknowledges support from the project VEGA 2/0004/16. Y.S. acknowledges the Major International Joint Research Project (11820101002) of NSFC, the Joint Research Fund in Astronomy (U1631242, U1731241) under the cooperative agreement between NSFC and CAS, the CAS Strategic Pioneer Program on Space Science, (grant No. XDA15052200, XDA15320300, XDA15320301), the Thousand Young Talents Plan, and the Jiangsu ``Double Innovation Plan". A.H. acknowledges Dr. Ewan C. Dickson for the proofread of this manuscript. \sdo\ data are courtesy of the NASA/\sdo\ and HMI science team. \rhessi\ is a NASA Small Explorer Mission. \goes\ is a joint effort of NASA and the National Oceanic and Atmospheric Administration (NOAA).


\bibliography{bibliography}   

\end{document}